\documentstyle[12pt,equation]{article}
\begin{document}
\newcommand{\ycut}{\mbox{$y_{\rm cut}$}}
\newcommand{\ph}{\mbox{$P_{\rm had}$}}
\newcommand{\phsq}{\mbox{$P^2_{\rm had}$}}
\newcommand{\delphi}{\mbox{$\Delta \phi$}}
\newcommand{\csoft}{\mbox{$\chi_{\gamma \gamma}^{\rm soft}$}}
\newcommand{\ppbar}{\mbox{$p \overline{p} $}}
\newcommand{\ccbar}{\mbox{$c \overline{c} $}}
\newcommand{\aem}{\mbox{$\alpha_{{\rm em}}$}}
\newcommand{\shat}{\mbox{$\hat{s}$}}
\newcommand{\sigtot}{\mbox{$\sigma_{\rm tot}(\gamma \gamma \rightarrow
{\rm hadrons})$}}
\newcommand{\sigjet}{\mbox{$\sigma_{\gamma\gamma}^{\rm jet}$}}
\newcommand{\sighat}{\mbox{$\hat{\sigma}$}}
\newcommand{\rs}{\mbox{$\sqrt{s}$}}
\newcommand{\pT}{\mbox{$p_T$}}
\newcommand{\wgg}{\mbox{$W_{\gamma \gamma}$}}
\newcommand{\fge}{\mbox{$f_{\gamma|e}$}}
\newcommand{\ptmin}{\mbox{$p_{T,{\rm min}}$}}
\newcommand{\qvec}{\mbox{$\vec{q}^{\gamma}$}}
\newcommand{\xq}{\mbox{$(x,Q^2)$}}
\newcommand{\qig}{\mbox{$q_i^{\gamma}$}}
\newcommand{\Gg}{\mbox{$G^{\gamma}$}}
\newcommand{\qqbar}{\mbox{$q \overline{q}$}}
\newcommand{\epem}{\mbox{$e^+e^-$}}
\newcommand{\gaga}{\mbox{$\gamma\gamma$}}
\newcommand{\be}{\begin{equation}}
\newcommand{\ee}{\end{equation}}
\newcommand{\een}{\end{subequations}}
\newcommand{\ben}{\begin{subequations}}
\newcommand{\beq}{\begin{eqalignno}}
\newcommand{\eeq}{\end{eqalignno}}
\renewcommand{\thefootnote}{\fnsymbol{footnote} }

\pagestyle{empty}
\begin{flushright}
MAD--PH--867\\
January 1995\\
\end{flushright}
\vspace{1cm}
\begin{center}
{\Large \bf Inclusive Two--Photon Reactions at TRISTAN}\footnote{To appear in
the Proceedings of the {\it Fourth Workshop on TRISTAN Physics at High
Luminosities}, KEK, Tsukuba, November 1994.}
\\
\vspace{5mm}
Manuel Drees\footnote{Heisenberg Fellow}\\
{\it Univ. of Wisconsin, Dept. of Physics, Madison, WI 53706, USA} \\

\end{center}

\vspace{2cm}
\begin{abstract}
After briefly reviewing past accomplishments of TRISTAN experiments in the
field of inclusive two--photon reactions, I discuss open problems in the Monte
Carlo simulation of such reactions. The main emphasis is on multiple
scattering, i.e. events where at least two pairs of partons scatter within
the same \gaga\ collision to form at least four (mini)jets. The cross section
for such events might just be observable at TRISTAN. While theoretical
arguments for the existence of such events are strong, they have not yet been
directly observed experimentally, thereby potentially opening a new
opportunity for TRISTAN experiments.

\end{abstract}
\vspace{1.5cm}
\clearpage
\setcounter{page}{1}
\pagestyle{plain}
\section*{1. Introduction}
In this talk I will attempt to cover some topics relevant for the
understanding of inclusive (mini)jet production in quasi--real two--photon
production at TRISTAN and elsewhere. In my opinion the study of such reactions
is interesting for at least two reasons. First, it increases our knowledge of
the perturbative structure of the photon, described by the parton densities
inside the photon \cite{1}. This is necessary to improve our ability to
predict (background) cross sections at higher energy $ep$ \cite{2} and \epem\
\cite{3} colliders. An accurate determination of these parton densities should
also allow to test predictions \cite{1,4} for these densities based on certain
dynamical assumptions.

This latter point is connected to the second main motivation for studying
two--photon reactions: They are an excellent testing ground for our
understanding of those aspects of semi--hard and non--perturbative QCD that
are relevant for collider phenomenology \cite{5}. On the one hand \gaga\
collisions ``ought" to be more easily treatable than fully hadronic
collisions, since intuitively a photon should be a simpler object than a
proton. On the other hand, the current description \cite{6} of
non--diffractive \gaga\ reactions in terms of direct, single resolved and
double resolved contributions, or the even more complicated classification
scheme of ref.\cite{5}, makes the complete understanding of \gaga\ reactions
appear considerably more challenging than that of \ppbar\ or $pp$ scattering.

In recent years TRISTAN experiments have contributed greatly to our
understanding of inclusive two--photon reactions. In particular, in 1991 the
AMY collaboration for the first time established \cite{7} the existence of
resolved photon contributions. This was also the first experiment that
succeeded in describing their data with a QCD--based Monte Carlo program;
previous codes had not included hard resolved photon contributions, and had
consequently not been able to reproduce PEP and PETRA data. A little later,
the TOPAZ collaboration presented \cite{8} a first measurement of cross
sections for the production of fully reconstructed jets, using a cone
algorithm. This is a great improvement over the previously used definition of
jets as ``thrust hemispheres", which obscured the relation between partons and
jets. At the same time, and approximately simultaneous with HERA experiments
\cite{9,2}, TOPAZ directly observed the spectator or remnant jets
characteristic for resolved photon processes. Most recently, TOPAZ has begun
to use \cite{10} this ability to tag remnant jets to disentangle direct and
resolved photon contributions to inclusive charm and $K^0$ production.
Finally, it should be mentioned that the LEP experiments ALEPH \cite{11} and
DELPHI \cite{12} have published first results on inclusive no--tag two--photon
reactions, and VENUS is also entering the fray \cite{13}.

On the theoretical side, progress has been made in the calculation of
next--to--leading order (NLO) corrections. Full NLO calculations for
single--jet inclusive cross sections (for massless partons) \cite{14}, and for
inclusive charm production \cite{15}, are available. Improved estimates for
the photon flux factors relevant for resolved photon contributions have been
presented in \cite{16}. Finally, the JETSET/PYTHIA program package has been
extended to include all classes of inclusive \gaga\ reactions \cite{5}.

In spite of this progress, open problems remain. The for practical purposes
most urgent problem is probably the lack of a reliable ``standard" MC code,
which is necessary to link parton--level calculations \cite{6,14} to
measured quantities. At present all four experimental groups active in this
field use their own MC generators. One measure of the differences between
these codes is the value of the cut--off parameter \ptmin\ (the minimal
allowed partonic transverse momentum in ``hard" resolved photon collisions)
determined from their respective data, even when assuming the same parton
densities in the photon: While AMY and TOPAZ now both quote \cite{8} values
around 2.0 GeV for the DG parametrization \cite{17}, DELPHI finds \cite{12}
a value of about 1.45 GeV, while ALEPH gives \cite{11} a value as large as
2.5 GeV. These groups all use different trigger criteria, and also use
different methods to determine the optimal value of \ptmin. Nevertheless this
quite substantial discrepancy\footnote{Note that the integrated double
resolved contribution to the jet cross section for $p_T > \ptmin$ scales
approximately like $p_{T, {\rm min}}^{-3.5}$; increasing \ptmin\ from 1.45 to
2.5 GeV therefore decreases this contribution by a factor of approximately
6.5!} indicates that (some) current generators are not yet complete.

In fact, to the best of my knowledge none of the generators used by
experiments to date includes initial state radiation (ISR). The implementation
if ISR in standard PYTHIA apparently leads to very poor agreement with the
data \cite{11}. I have argued elsewhere \cite{18} that ISR in resolved photon
interactions should be cut off sooner than in \ppbar\ scattering, at least
whenever the ``pointlike" component of the photon structure functions is
involved. I believe the program developed in ref.\cite{5} allows to do so at
least as an option. I would like to stress here that some amount of ISR {\em
has} to exist; switching it off completely is certainly an
over--simplification. One possibility to investigate this experimentally is to
study the opening angle in the transverse plane \delphi\ between the two jets
in two--jet events. On the parton level (and in the absence of multiple
interactions, to be discussed below), $\delphi = 180^{\circ}$ unless final and
initial state radiation are included. The more \delphi\ deviates from
$180^{\circ}$, the more ISR is present.\footnote{Final state radiation can
also give $\delphi < 180^{\circ}$. However, I see no reason why standard
prescriptions for FSR should fail for resolved photon events; FSR can
therefore be subtracted, or included, using standard MC codes.}

Another potentially important ingredient of a successful MC code is the
treatment of multiple partonic scatterings in the {\em same} \gaga\
scattering; see Fig.~1. For example, PYTHIA makes use of multiple scattering
to reproduce features of the ``underlying event" at \ppbar\ colliders
\cite{19}. This effect is included in the treatment of ref.\cite{5}, but has
otherwise not been discussed in any detail in the context of two--photon
reactions. I therefore decided to use this Contribution to provide a first
quantitative estimate of multiple scattering event rates in \gaga\ collisions.
To that end, I briefly describe in Sec.~2 the relation between multiple
scattering and calculations of the total \gaga\ cross section at high energies
in the minijet picture. Sec.~3 contains estimates for multiple scattering
event rates at TRISTAN, and Sec.~4 is devoted to a brief summary and
conclusions.

\setcounter{footnote}{0}
\section*{2. Multiple scattering and $\sigma_{\rm tot}(\gamma \gamma
\rightarrow$ hadrons)}
In leading order QCD, the inclusive cross section for the production of (at
least) one jet pair with $p_T \geq \ptmin$ in \gaga\ collisions is given by
the well--known expression
\be \label{e1}
\sigjet = \sum_{i,j,k,l} \int_{x_{\rm min}}^1 dx_1 \int_{\frac{x_{\rm min}}
{x_1}}^1 dx_2 f_{i|\gamma} (x_1) f_{j|\gamma}(x_2)
\int_{\ptmin}^{\sqrt{\hat{s}}/2} \frac {d \hat{\sigma}_{ij \rightarrow kl}
(\hat{s})} {dp_T} dp_T,
\ee
with $x_{\rm min} = 4 p^2_{T,{\rm min}}/s$, $s$ being the squared \gaga\
centre--of--mass energy, and $\hat{s} = x_1 x_2 s$. $f_{i|\gamma}$ is the
density of parton $i$ in the photon, and the $\sighat_{ij \rightarrow kl}$
are the hard QCD $2 \rightarrow 2$ scattering cross sections \cite{20}.
Eq.(\ref{e1}) is straightforward to evaluate numerically for given parton
distribution functions $f_{i|\gamma}$ and given \ptmin. For example, using
the DG parametrization \cite{17} one finds approximately \cite{21}
\be \label{e2}
\sigjet({\rm DG}) \simeq 270 \ {\rm nb} \ \left( \frac {\sqrt{s}}
{50 \ {\rm Gev}} \right)^{1.4} \left( \frac {1.6 \ {\rm GeV}}
{\ptmin} \right)^{3.6},
\ee
where I have used $Q^2 = \hat{s}/4$ as scale in $\alpha_S$ and $f_{i|\gamma}$.

Notice that the cross section (\ref{e2}) increases like a power of the cms
energy $\sqrt{s}$. At sufficiently high energy it will therefore exceed the
usual VDM estimate \cite{22} for the {\em total} cross section for $\gaga
\rightarrow$ hadrons,
\be \label{e3}
\sigma_{\gamma\gamma}^{\rm VDM} \simeq 250 \ {\rm nb} + \frac {300 \ {\rm nb
\ GeV}} {\sqrt{s}}.
\ee
One possibility is, of course, that the total hadronic cross section at high
ernergies is indeed much larger than the VDM estimate (\ref{e3}). However,
given the rather modest increase of the total $\gamma p$ cross section as
measured at HERA \cite{23}, which can be described quite well \cite{24} by
the ``universal" asymptotic $s^{0.08}$ behaviour also found in total \ppbar\
and $pp$ cross sections, a rapid increase of \sigtot\ now seems implausible.

It is important to notice here that eqs.(\ref{e1},\ref{e2}) refer to an
{\em inclusive} cross section, which by definition contains a factor of the
average jet pair multiplicity:
\be \label{e4}
\sigjet = \langle n_{\rm jet \ pair} \rangle \cdot \sigtot.
\ee
This means that whenever eq.(\ref{e2}) exceeds the total hadronic cross
section, there must be on average more than one jet pair per \gaga\ collision.
Since we are still working in leading order in QCD, the only possibiltiy to
produce additional jet pairs is to have several parton--parton scatterings in
the same \gaga\ scattering.\footnote{Higher order QCD corrections will not
change this conclusion, since here the cross section for the production of
many jets is suppressed by powers of $\alpha_s$; this suppression is not
compensated by any enhancement factors, except in the relatively rare cases
where $\hat{s} \gg 4 p_T^2$. In contrast, the comparison of eqs.(\ref{e2})
and (\ref{e4}) shows that $\langle n_{\rm jet \ pair} \rangle$ must grow like
some power of the cms energy, if $\sigma_{\rm tot}$ grows only slowly with
energy.} Hence there is very strong evidence from perturbative QCD that
multiple partonic interactions within one \gaga\ event {\em must} occur at
high energies.\footnote{The only way around this conclusion is to make
$p_{T,{\rm min}}$ grow with $\sqrt{s}$. However, intuitively $p_{T,{\rm min}}$
should describe a cut--off due to confinement effects. It is difficult to
understand why this should depend on the energy.}

The simplest quantitative estimates of multiple interaction rates are based on
the eikonal formalism \cite{25,19}. One writes the total interaction cross
section as
\be \label{e5}
\sigtot = \phsq \int d^2 b \left[ 1 - e^{-\chi(s) A(b) / P^2_{\rm had}}
\right].
\ee
Here $\vec{b}$ is the (two--dimensional) impact parameter, $A(b)$ describes
the distribution of scatter centers (i.e., partons) in the transverse plane,
and the dynamical information about the individual scattering processes is
contained in the eikonal $\chi(s)$. The quantity \ph, first introduced in
ref.\cite{26}, describes the probability for a photon to go into a hadronic
state, and is thus of order \aem.

It is customary to split the eikonal into a soft (nonperturbative) part
\csoft\ and the perturbative contribution, which is nothing but the minijet
contribution (\ref{e1}):
\be \label{e6}
\chi(s) = \csoft(s) + \sigjet(s).
\ee
The physical meaning of the ansatz (\ref{e5}) then becomes more transparent
when it is re--written as \cite{25}
\beq \label{e7}
\sigtot &= \phsq \int d^2b \ e^{-\sigma_{\gamma\gamma}^{\rm jet}(s) A(b) /
P^2_{\rm had} } \left[ e^{\sigma_{\gamma\gamma}^{\rm jet}(s) A(b) /
P^2_{\rm had} } - e^{-\chi^{\rm soft}_{\gamma\gamma}(s) A(b) / P^2_{\rm had} }
\right] \\
&= \phsq \int d^2b \ e^{-\sigma_{\gamma\gamma}^{\rm jet}(s) A(b) /
P^2_{\rm had} } \left[ \sum_{n=0}^{\infty} \frac {1}{n!} \left( \frac
{\sigma_{\gamma\gamma}^{\rm jet}(s) A(b)}{P^2_{\rm had}} \right)^n
 - e^{-\chi^{\rm soft}_{\gamma\gamma}(s) A(b) / P^2_{\rm had} } \right].
\nonumber
\eeq
Notice that each term in the sum, when multiplied with the exponential in
front of the square brackets, is equivalent to the Poisson probability to have
$n$ {\em independent} hard scatters at impact parameter $b$, the average
number of scatters being $\sigma_{\gamma\gamma}^{\rm jet}(s) A(b) /
P^2_{\rm had}$. Note that \sigjet\ is of order $\alpha^2_{\rm em}$, since the
parton densities $f_{i|q}$ in eq.(\ref{e1}) are of order \aem. The presence
of the factor \phsq\ in the denominator then ensures that the probability for
additional hard scatters is {\em not} suppressed by additional powers of \aem;
this is reasonable, since the transition of the incident photons into hadronic
states only has to occur once, independent of the number of hard scatters.
Furthermore, the fact that we obtain a Poisson distribution for the number of
scatters at fixed impact parameter means that we have {\em assumed} that these
reactions occur {\em independently} of each other; we will see later that this
assumption might be questionable in case of \gaga\ collisions.

The fact that the perturbative contribution to the eikonal $\chi$ of
eq.(\ref{e6}) should indeed be the jet cross section (\ref{e1}) can be seen by
computing the {\em inclusive} cross section for the production of (at least)
$k$ jet pairs from eq.(\ref{e7}). To that end, one simply has to include all
terms with $n \geq k$ in the sum, and multiply them with the combinatorics
factor $\left( \begin{array}{c} n \\ k \end{array} \right)$ to pick $k$ jet
pairs out of a total of $n$ pairs:
\beq \label{e8}
\sigma( \geq k \ {\rm jet \ pairs}) &= \phsq \int d^2b \
e^{-\sigma_{\gamma\gamma}^{\rm jet}(s) A(b) / P^2_{\rm had} }
\sum_{n=k}^{\infty} \frac {1}{n!} \left( \begin{array}{c} n \\ k \end{array}
 \right) \left( \frac {\sigma_{\gamma\gamma}^{\rm jet}(s) A(b)}
{P^2_{\rm had}} \right)^n \nonumber \\
 &= \phsq \frac {1} {k!} \int d^2b \
e^{-\sigma_{\gamma\gamma}^{\rm jet}(s) A(b) / P^2_{\rm had} }
\sum_{n=k}^{\infty} \frac {1}{(n-k)!}
\left( \frac {\sigma_{\gamma\gamma}^{\rm jet}(s) A(b)}
{P^2_{\rm had}} \right)^{(n-k)+k} \nonumber \\
&= \phsq \frac{1}{k!} \int d^2b \left( \frac
{\sigma_{\gamma\gamma}^{\rm jet}(s) A(b)} {P^2_{\rm had}} \right)^k.
\eeq
This just gives \sigjet\ for $k=1$, since $\int d^2 A(b) = 1$ by definition.
Of more interest for us is the cross section for having at least a second
independent partonic collision within the same \gaga\ event, which is given
by eq.(\ref{e8}) with $k=2$:
\beq \label{e9}
\sigma(\geq 2 \ {\rm jet \ pairs}) &= \frac{1}{2} \left[ \sigjet(s) \right]^2
\int d^2b \frac {A(b)^2} {P^2_{\rm had}} \nonumber \\
&\equiv \left[ \sigjet(s) \right]^2 / \sigma_0,
\eeq
where I have introduced
\be \label{e10}
\sigma_0 = \frac {2 \phsq }{ \int d^2b A(b)^2}.
\ee

Obviously the quantity $\sigma_0$ will play a crucial role in estimates of
multiple scattering event rates. Intuitively it is something like the
probability \phsq\ for the two incident photons to go into hadronic systems,
multiplied with the geometrical cross section of these systems. Clearly the
same quantities that determine the numerical value of $\sigma_0$, i.e.
\ph\ and $A(b)$, also enter the prediction (\ref{e5}) for the total hadronic
cross section at high energies. In order to make this connection more
quantitative let us consider the simple Gaussian ansatz
\be \label{e11}
A(b) = \frac {1}{\pi b_0^2} e^{-b^2/b_0^2}
\ee
for the transverse distribution of partons in the photon. It is then quite
easy to see that all physical quantities will only depend on the product
$\ph \cdot b_0$. In particular, this ansatz gives
\be \label{e12}
\sigma_0 = 4 \pi b_0^2 \phsq.
\ee
We can therefore somewhat arbitrarily fix $\ph=1/200$ and explore the model
dependence by varying $b_0$, which characterizes the transverse size of the
hadronic system.

Fig.~2 shows predictions for the energy dependence of the total \gaga\ cross
section as computed from eq.(\ref{e5}) using the DG parametrization \cite{17}
with $\ptmin=1.6$ GeV and three different values of $b_0$. For this
calculation I have assumed
\be \label{e13}
\csoft(s) = \chi_0 + \frac {\chi_1}{\sqrt{s}},
\ee
as indicated by the VDM prediction (\ref{e3}); numerically, $\chi_0 = 0.375$
(0.5, 1.0) $\mu$b and $\chi_1=1.1$ (2.5, 12.5) $\mu$b$\cdot$GeV for $b_0=2.3$
(1.9, 1.5) GeV$^{-1}$. The largest value of $b_0$ shown, 2.3 GeV$^{-1} = 0.46$
fm, would lead to a substantial increase of the total \gaga\ cross section
already at $\sqrt{s}=100$ GeV, while the smallest choice, $b_0=1.5 \ {\rm
GeV}^{-1} = 0.3$ fm, leads to a very slow rise of this cross section, in rough
agreement with the universal $s^{0.08}$ behaviour postulated by Donnachie and
Landshoff \cite{24}. It is crucial to keep in mind that smaller values of
$b_0$ give {\em smaller} total cross sections (\ref{e5}) at high energies, but
also smaller values for $\sigma_0$ (\ref{e12}) and hence {\em larger} rates
for multiple scattering events (\ref{e9}); the three values of $b_0$ shown in
Fig.~2 correspond to $\sigma_0 = 645$, 440, and 275 nb, respectively. This
inverse relation between the total cross section and the rate of multiple
scattering events is a direct consequence of eq.(\ref{e4}), i.e. it is
independent of the specific ansatz for $A(b)$, or even of the eikonal ansatz
(\ref{e5}) for \sigtot. I repeat, the {\em smaller} the total cross section at
high energies, the {\em larger} the rate for multiple scattering events at a
given energy.

\setcounter{footnote}{0}
\section*{3. Multiple Scattering Rates at TRISTAN}
In order to compute rates at \epem\ colliders for events containing multiple
partonic scatters within one \gaga\ collision, one has to convolute the \gaga\
cross section with photon flux factors in the standard way. In particular, the
cross section (\ref{e9}) for the production of at least two jet pairs becomes:
\beq \label{e14}
\sigma_{e^+e^-}(\geq 2 \ {\rm jet \ pairs}) &= \frac {1} {\sigma_0}
\sum_{\rm partons} \int_{z_{1,{\rm min}}}^1 d z_1 \fge(z_1)
\int_{z_{2,{\rm min}}}^1 d z_2 \fge(z_2)
\int_{x_{1,{\rm min}}}^1 d x_1 f_{i|\gamma}(x_1) \nonumber \\
&\cdot \int_{x_{2,{\rm min}}}^1 d x_2 f_{j|\gamma}(x_2)
\int_{\ptmin}^{\sqrt{\hat{s}}/2} d p_T \frac
{d \hat{\sigma}_{ij \rightarrow kl}(\hat{s})} {d p_T} \nonumber\\
&\cdot \int_{x_{1,{\rm min}}}^{1-x_1} d x'_1 f_{i'|\gamma}(x'_1)
\int_{x'_{2,{\rm min}}}^{1-x_2} d x'_2 f_{j'|\gamma}(x'_2)
\int_{\ptmin}^{\sqrt{\hat{s'}}/2} d p'_T \frac
{d \hat{\sigma}_{i'j' \rightarrow k'l'}(\hat{s'})} {d p'_T}.
\eeq
Here, $z_{1,{\rm min}} = 4 p_T^2/s, \ z_{2,{\rm min}} = z_{1,{\rm min}}/z_1, \
x_{1,{\rm min}} = z_{2,{\rm min}}/z_2, \ x_{2,{\rm min}} = x_{1,{\rm min}} /
x_1, \ x'_{2,{\rm min}} = x_{1,{\rm min}}/x'_1, \ \hat{s} = z_1 z_2 x_1 x_2 s$
and $\hat{s}' = z_1 z_2 x'_1 x'_2 s$. Notice that even for given $z_1$ and
$z_2$ the expression (\ref{e14}) does not factorize into two independent cross
sections, as was indicated in eq.(\ref{e9}). The reason is that I modified the
upper boundaries of the integrations over $x'_1$ and $x'_2$ in order to
enforce energy--momentum conservation, i.e. $x_1 + x'_1 \leq 1$ and
$x_2 + x'_2 \leq 1$. A slightly different method to do this has been used in
ref.\cite{19}. However, this requirement is much more important for \gaga\
collisions than for \ppbar\ collisions, since the quark densities inside the
photon are much harder than those inside the proton, i.e. they remain sizable
for $x$ quite close to 1. I will come back to this point later.

In principle the sum in eq.(\ref{e14}) contains 64 independent terms
(combinations of parton densities and hard subprocess cross sections
$\hat{\sigma}$). The calculation can be greatly simplified by using the
observation of ref.\cite{27} that the sums over parton species can be
treated approximately by introducing the effective parton density
\be \label{e15}
f_{{\rm eff}|\gamma}(x) = \frac{9}{4} f_{G|\gamma}(x) + 2 \sum_i
f_{q_i|\gamma}(x),
\ee
where the factor of two takes care of anti--quarks. In the same approximation
all hard scattering cross sections are replaced by the cross section for
the elastic scattering of two different quarks, $\hat{\sigma}(q q' \rightarrow
q q')$. The sum in eq.(\ref{e14}) then collapses to a single term. I checked
that this approximation reproduces the ``exact" (leading order) prediction for
the single jet pair inclusive cross section, eq.(\ref{e1}), to better than
10\%, which is considerably smaller than the overall theoretical uncertainty
of this estimate of multiple scattering rates.

Eq.(\ref{e14}) describes the production of (at least) four jets. However, not
all of them have to fall within the angular region covered by a given
detector, i.e. have rapidity $|y| \leq \ycut$. At the parton level, the jet
rapidities are given by the usual relations
\be \label{e16}
y_{1,2} = \log \left[ \frac {x_1 z_1} {x_T} \left( 1 \pm \sqrt{ 1 -
\frac {x_T^2} {z_1 z_2 x_1 x_2} } \right) \right],
\ee
where $x_T = 2 p_T / \sqrt{\hat{s}}$; the rapidities $y'_{1,2}$ of the second
pair of jets are given by eq.(\ref{e16}) with $(x_1,x_2,x_T) \rightarrow
(x'_1, x'_2, x'_T)$. For a given \ycut\ one can then compute five independent
cross sections from eq.(\ref{e14}), depending on the number of jets that
satisfy $|y_i| \leq \ycut$, which I denote by $\sigma_{n4} \ (1 \leq n \leq
4)$; moreover, for $\sigma_{24}$ I distinguish contributions where both
detected jets come from the same partonic scattering $(\sigma_{24a})$ from
those where both jet pairs contribute one jet each $(\sigma_{24b})$. Notice
that these $\sigma_{n4}$ are (approximately) {\em exclusive} cross sections,
i.e. $\sigma_{14}$ is the cross section for having {\em exactly one} jet
with $|y| \leq \ycut$, and so on.\footnote{These cross sections are only
approximately exclusive since in the derivation of eq.(\ref{e8}), which led
to eq.(\ref{e14}), all terms with $n \geq 2$ where included. In other words,
eq.(\ref{e14}) includes contributions with three, four, \dots, independent
scatters, some of which might produce additional jets in the acceptance
region. However, we will see below that, at least at TRISTAN energies, the
cross section for the simultaneous production of at least two independent jet
pairs at {\em any} rapidity is almost certainly substantially smaller than
that for the production of a single jet pair. This indicates that for
two--photon cms energies of relevance for TRISTAN, the sum in eq.(\ref{e8}) is
still dominated by its first term, so that the rate for producing at least
two jet pairs is very close to that for producing exactly two jet pairs.}

Fig.~3 shows the dependence of these five cross sections on \ptmin, where I
have taken $\sqrt{s} = 58$ GeV, $\ycut=1$, and $\sigma_0=300$ nb
(corresponding to $b_0 = 1.6 \ {\rm GeV}^{-1}$; see Fig.~2). I have included
the anti--tagging condition $\theta_e \leq 5^{\circ}$ for the outgoing
electron and positron when computing the photon flux functions $f_{\gamma|e}$,
and have estimated the suppression due to the virtuality of the photon as
described in ref.\cite{16}. For comparison this figure also shows single pair
inclusive cross sections, split into contributions where exactly one
$(\sigma_{12})$ or both $(\sigma_{22})$ jets pass the rapidity cut. These
contributions, represented by the dotted curves, are nothing but the standard
(LO) predictions \cite{6} for double resolved jet production at TRISTAN. If
single resolved and direct contributions were added, the dotted curves would
have to be pushed up by a factor of 3 to 5. However, recent results on
spectator jet tagging \cite{10} indicate that this might not be
necessary.\footnote{It should be clear that multiple interactions can only
occur in double resolved \gaga\ reactions, since a pointlike (direct) photon
is ``used up" after a single interaction.}

Evidently rates for events with multiple hard scattering are not very large at
TRISTAN, not even for $\ptmin=1.6$ GeV, the smallest value shown. Moreover, a
fraction of these events, given by $\sigma_{14}$ and $\sigma_{24a}$, have the
same partonic final state within the given rapidity range as the ``standard"
contributions $\sigma_{12}$ and $\sigma_{22}$, respectively. Indeed, those
parts of $\sigma_{14}$ and $\sigma_{24a}$ where the detected jet(s) come(s)
from the ``unprimed" (first) partonic collision in eq.(\ref{e14}) are already
included in $\sigma_{12}$ and $\sigma_{22}$. Recall that these are {\em
inclusive} cross sections as far as additional partonic scatterings are
concerned; e.g., $\sigma_{22}$ is the cross section for having both jets
produced in the {\em first} partonic scattering inside the acceptance region,
{\em independent} of whether or not additional hard scatters occur in the same
event.\footnote{Similarly, $\sigma_{34}$ is included partly in $\sigma_{12}$
and partly in $\sigma_{22}$, and $\sigma_{44}$ is included entirely in
$\sigma_{22}$.}

Assuming that nothing at all is known about particle flows in the region
$|y| > \ycut$ (other than perhaps the existence of spectator jets),
information about multiple scattering events must therefore come from the
contributions described by $\sigma_{24b}, \ \sigma_{34}$ and $\sigma_{44}$.
Clearly the most distinctive signature would be the detection of all four
jets. Except for (small) corrections due to initial and final state radiation,
one expects these four jets to occur in two back--to--back pairs with equal
and opposite transverse momentum, i.e. $\vec{p_T}(j1) \simeq - \vec{p_T}(j2)$
and  $\vec{p_T}(j3) \simeq - \vec{p_T}(j4)$. Moreover, and in sharp contrast
to multi--jet final states produced by higher order QCD processes from a
single parton pair, the angular distribution in the transverse plane between
these two jet pairs should be flat, i.e. the pairs should be uncorrelated.
Unfortunately Fig.~3 shows that, given TRISTAN's integrated luminosity of
a few hundred pb$^{-1}$, each group will at best find a handful of such events
with $\ycut=1$, unless $\sigma_0$ is substantially smaller than 300 nb; notice
that no allowance for finite jet reconstruction efficiencies has yet been
made.

On the other hand, the rate of three--jet events that are due to multiple
interactions, described by $\sigma_{34}$, might well be detectable. Up to
ISR and FSR effects, these events should have the configuration $\vec{p_T}(j1)
\simeq - \vec{p_T}(j2)$, with a flat $\Delta \phi$ distribution of the third
jet with respect to the other two. Clearly a full MC analysis will be
necessary to decide whether these properties make such events sufficiently
distinguishable from ``ordinary" QCD 3--jet events. A good understanding of
initial and final state radiation will certainly be crucial for this study.

Finally, the cross sections for events with at least two independent partonic
scatters clearly drop much faster with increasing \ptmin\ than the single jet
pair inclusive cross sections do. This is not surprising. Eq.(\ref{e14}) shows
that additional hard scatters mean additional factors of $d \hat{\sigma} / d
p_T$ (which decreases rapidly with increasing $p_T$) and additional factors of
$f_{{\rm eff}|\gamma}(x_i)$ (which decrease with increasing $x_i$; recall that
the lower bound on the $x_i$ scales like $p^2_{T,{\rm min}}$). Clearly finding
any evidence for multiple scattering at TRISTAN will be hopeless if one
requires $\ptmin \geq 2.5$ GeV or so for {\em all} detected jets. This
indicates that most present analyses \cite{8}, which ignored multiple
scattering and focussed on relatively large $p_T$, will remain unaffected. It
also means that such events with comparatively high $p_T$ can be used to study
initial and final state radiation without having to worry about multiple
interaction effects, as described in Sec.~1. Once ISR and FSR are understood,
one can go back to smaller $p_T$ and look for the contribution $\sigma_{34}$
as described above. One might then even be able to find evidence for the
contribution $\sigma_{24b}$, where one has two jets which are usually neither
back--to--back nor have equal $| \vec{p_T}|$; these events should show up in
the tails of the $\Delta \phi$ distribution discussed in Sec.~1.

Fig.~4 shows the dependence of the various cross sections on the rapidity cut,
for $\ptmin=1.6$ GeV. In the limit $\ycut \rightarrow \infty$ only
$\sigma_{22}$ and $\sigma_{44}$ remain finite since all produced jets are now
detected; the fact that $\sigma_{22} > 5 \sigma_{44}$ even in this limit, and
for the small value of \ptmin\ chosen, once again indicates that multiple
scattering events are indeed quite rare at TRISTAN. However, if the rapidity
coverage could be extended to 1.5 or even 2, the chances of detecting
four--jet events due to independent partonic scatters would improve
dramatically compared to the case with $\ycut=1$ shown in Fig.~3. Of course,
it is no longer possible to modify TRISTAN detectors; however, it might be
possible to reconstruct jets using calorimetric information only. Although this
will presumably come at the cost of substantially increased errors on the
$\vec{p_T}$ of jets outside the core region of the detector, it could still
greatly facilitate the study of multiple interactions.

It might be appropriate to briefly discuss some of the uncertainties of my
estimates of multiple interaction rates here. As emphasized in Sec.~2, the
overall rate scales like $1/\sigma_0$, which in turn largely determines the
high--energy behaviour of \sigtot. The value $\sigma_0=300$ nb used for my
numerical estimates corresponds to $b_0=1.6$ GeV$^{-1}$, and thus to a rather
modest growth of the total cross section, as shown in Fig.~2. Of course, the
multiple interaction rate also depends on the same quantities that determine
the usual (leading order) jet cross sections, i.e. the parton distribution
functions, the value of the QCD scale parameter $\Lambda$, and the momentum
scale to be used in $\alpha_S$ and in the parton densities; in fact,
eq.(\ref{e14}) shows that the cross section for multiple interactions depends
twice as strongly on these parameters as ordinary (inclusive) jet cross
sections do.

Finally, there is the question of how reliable the eikonal ansatz (\ref{e5})
is, on which my estimates are based. As already mentioned earlier, this ansatz
only holds {\em if} multiple interactions occur independently of each other.
This might be a good approximation for \ppbar\ collisions, where such an
ansatz has been tested most extensively \cite{19}, since here one deals with
real hadrons that do certainly contain several partons, even though the exact
relation between partons and constituent quarks might be quite complicated.
In contrast, in the final analysis the entire parton content of the photon can
be traced back to the $\gamma q \overline{q}$ vertex. One would therefore
expect dynamical correlations to exist between different partons ``in" the
photon. These will probably be stronger at large Bjorken--$x$, since partons
at large $x$ cannot be far removed from the primary vertex, i.e. not many
parton splittings can have occured, starting from the original  $\gamma q
\overline{q}$ vertex, to produce a parton at large $x$. One might therefore
expect the assumption of independent scatters to work better if one sticks to
relatively soft partons. Unfortunately, requiring $x_1 + x'_1 \leq 0.5$ and
$x_2 + x'_2 \leq 0.5$ in eq.(\ref{e14}) reduces the total multiple interaction
rate by more than a factor of 4, making them very difficult to detect
experimentally.\footnote{For $\ycut=1$, this restriction of the Bjorken$-x$
variables only reduces $\sigma_{44}$ by about a factor of two, since events
with at least one large Bjorken$-x$ in the initial state tend to be more
strongly boosted, i.e. to produce jets at large rapidities. I should also
mention that in ref.\cite{5} the eikonal ansatz is used only for the
``hadronic" component of the photon structure function. There the contribution
from the ``pointlike" or ``anomalous" component of the $f_{i|\gamma}$ is
regularized by increasing the cutoff \ptmin\ linearly with the cms energy. I
find this treatment not very satisfying. For one thing, the sharp separation
in ``hadronic" and ``pointlike" pieces is clearly an over--simplification.
Also, as mentioned earlier, I find it hard to understand why \ptmin\ should
depend on the beam energy. Nevertheless the generator of ref.\cite{5} is the
only existing two--photon MC code that treats multiple interactions at all.}

As mentioned earlier, the ansatz (\ref{e5}) (with $\ph \equiv 1$) has been
used to describe \ppbar\ collisions. One can certainly reproduce the observed
behaviour of the total cross section in this way \cite{28}, but other
successful descriptions exist as well \cite{24}. In ref.\cite{19} independent
multiple interactions were used to describe details of the event structure
at \ppbar\ colliders, some of which are difficult to understand otherwise. One
example is the so--called pedestal effect, i.e. the observation that events
with a very hard interaction, say a pair of jets with $x_T > 0.1$, have a
higher multiplicity and transverse energy flow even away from these jets than
minimum bias events do. This is expected in the eikonal picture, since such
events are more likely to be very central, i.e. to have impact parameter $b$
close to zero, which enhances the chance for additional partonic interactions.
Furthermore, we heard at this meeting \cite{2} that HERA data also have some
features that might be explainable in terms of multiple interactions.
Nevertheless no {\em direct} evidence for the existence of events with
multiple partonic interactions has yet been found. In particular, no clear
signal for the production of two independent jet pairs (equivalent to the
contribution $\sigma_{44}$ discussed here) has yet been observed anywhere.
Therefore questions about the reliability of the eikonal ansatz remain even in
case of the proton.

Why might two--photon experiments succeed where \ppbar\ experiments at much
higher energy failed? There are two reasons to be optimistic, which are
actually related to each other. In spite of the hadronic nature of the photon,
which gives rise to spectator jets in resolved photon processes, \gaga\
collisions are considerably ``cleaner" than \ppbar\ collisions. This leads to
the possibility, demonstrated during this workshop \cite{29}, to reconstruct
jets with very low transverse momentum, below 2 GeV; this is inconceivable at
hadron colliders \cite{30}, and is probably impossible even at HERA \cite{31}.
Secondly, \gaga\ collisions are considerable more ``jetty" than \ppbar\
events, that is the fraction of events with identifiable jets is larger. This
can partly be explained by the hardness of the quark densities inside the
photon, which allows to funnel a large fraction of the photon's energy into a
single parton, thereby allowing jet production (from resolved photons!) at
relatively small two--photon cms energy \wgg. Moreover, the soft (non--jet)
cross section seems to be anomalously small for photons, or large for protons;
that is, the ratio of total \ppbar\ to \gaga\ cross sections exceeds the naive
expectation of $1/\alpha^2_{\rm em}$ by almost an order of magnitude. In other
words, effectively the normalization $1/\sigma_0$, and hence the rate for
multiple interactions, is about an order of magnitude smaller for \ppbar\
than for \gaga\ collisions.

However, experiments looking for multiple interactions in two--photon events
at \epem\ colliders do face one obstacle not encountered at hadron colliders:
The presence of \epem\ annihilation events makes it necessary to impose upper
limits on the energy and/or invariant mass of the observed hadronic system.
Recall that the cross section for multiple interactions grows very quickly
with \wgg; see eqs.(\ref{e9}) and (\ref{e2}) in Sec.~2. These cuts might
therefore reduce the signal significantly. This is demonstrated in Fig.~5,
which shows the normalized distributions in \wgg\ and in the summed energy of
all jets, for four--jet events. Here I have chosen $\ycut=2$, so that the sum
over energies should approximate the total energy deposition from the
high$-p_T$ partons in the calorimeter of a typical TRISTAN detector. Notice
that some parts of the spectator jets are usually also detected, which shifts
this distribution to the right. On the other hand, some parts of the spectator
jets will almost always be lost in the beam pipes, so that the measured \wgg\
distribution will be somewhat softer than the one shown in Fig.~5.
Nevertheless it should be clear from this plot that the typical selection
cuts, $W_{\rm vis} \leq 15$ (20) GeV for TOPAZ (AMY), will reduce the signal
for multiple interactions significantly. It might therefore be worthwhile to
try and relax this cut, or to replace it, e.g. by a cut on the {\em
transverse} energy in the event, which will be quite small for all \gaga\
events, including those with multiple interactions.

\section*{4. Summary and Conclusions}
TRISTAN experiments have contributed greatly to our understanding of
multi--hadron production in \gaga\ collisions, and thus of the hadronic
structure of the photon. Recent theoretical progress, especially concerning
NLO QCD calculations, should allow to fully exploit these measurements. The
major stumbling block at present seems to be the lack of a reliable event
generator that allows to estimate effects due to parton showering and
fragmentation. Among other things, this is crucial for comparing results from
different experiments, which have different trigger criteria, acceptance cuts
etc. As mentioned in Sec.~1, one weakness of existing generators is the (lack
of) treatment of initial state radiation; another, related, effect that is
specific for resolved photon interactions is the relatively large intrinsic
$k_T$ of partons in the photon \cite{32}.

The main focus of this contribution was on events with multiple partonic
interactions in the {\em same} \gaga\ collision, since they might offer
another great opportunity for TRISTAN experiments. Indeed, in my view the
proof of the existence of such events would be even more important than the
proof \cite{7} of the existence of resolved photon interactions. As I argued
in Sec.~2, multiple interactions are expected to occur in {\em all} hadronic
collisions $(\ppbar, \ \gamma p$ and \gaga) at sufficiently high energies, yet
they have never been observed directly. The rate for such events is intimately
linked to the behaviour of total hadronic cross sections at high energies, a
subject of much debate (without definite conclusion!) for about thirty years.
The numerical estimates of Sec.~3 indicate that detection of such events at
TRISTAN will not be easy, but might be possible. Not only the ``gold plated"
four--jet events, but also certain classes of three-- and even two--jet events
might be utilized for demonstrating the existence of multiple interactions. I
argued that it should be advantageous, and might be necessary, to extend
previous analyses of two--photon reactions by trying to reconstruct jets using
calorimetric information only (in order to increase the angular coverage,
which is crucial for four--jet events), and/or by relaxing the upper limits on
$E_{\rm vis}$ and $W_{\rm vis}$ in the definition of the two--photon event
sample.

As discussed in Sec.~3, the numerical estimates presented here are quite
uncertain. In my opinion this should serve as additional stimulus for
experimenters to try and find such events. After all, a large theoretical
uncertainty means that little is known, so all information is helpful,
including negative one. In particular, a meaningful lower bound on the
quantity $\sigma_0$ might indicate that the total \gaga\ cross section grows
faster at high energies than presently anticipated, which can have
ramifications for the planning of future experiments. Positive evidence for
multiple interactions might indicate that previous estimates \cite{3} of the
transverse energy flow in typical \gaga\ events at high energies have to be
revised upwards. To my mind much more important than these rather mundane
considerations of backgrounds at future colliders is the exciting possibility
to learn something about an aspect of strong interactions about which little
is known to date. I eagerly look forward to the first experimental study of
multiple partonic interactions in two--photon collisions, at TRISTAN or
elsewhere.

\noindent
\subsection*{Acknowledgements}
I thank the organizers for inviting me to give a talk at this workshop, which
got me thinking about multiple interactions in \gaga\ collisions. I also thank
K. Hagiwara, M.M Nojiri, Nyanko--sensei and the other members of the KEK
theory group for their hospitality while the first preliminary work along
these lines was done. This work was supported in part by the U.S. Department
of Energy under contract No. DE-AC02-76ER00881, by the Wisconsin Research
Committee with funds granted by the Wisconsin Alumni Research Foundation, as
well as by a grant from the Deutsche Forschungsgemeinschaft under the
Heisenberg program.

\clearpage
\section*{Figure Captions}
\renewcommand{\labelenumi}{Fig.\arabic{enumi}}
\begin{enumerate}

\item  
A Feynman diagram for a \gaga\ collision with two separate parton--parton
collisions. All combinations of partonic reactions ($qq, \ qG$ and $GG$)
contribute to the total rate for events with multiple interactions.

\vspace*{5mm}
\item  
The dependence of the total hadronic two--photon cross section on the \gaga\
cms energy \wgg, as calculated in the simple eikonal model of Sec.~2, for
three different values of $b_0$.

\vspace*{5mm}
\item  
The dependence of the multiple interaction cross section on the transverse
momentum cut--off \ptmin, as computed from eq.(\ref{e14}). The total cross
section for the production of at least four jets is split into five different
contributions, depending on the number of jets with $|y| \leq \ycut = 1$, as
described in the text. The double resolved contribution to the single jet pair
inclusive cross sections are shown for comparison by the dotted curves. All
cross sections have been computed using an anti--tag cut $\theta_e \leq
5^{\circ}$ for the outgoing electron and positron.

\vspace*{5mm}
\item   
The dependence of jet cross sections on the acceptance region \ycut. The
notation is as in Fig.~3.

\vspace*{5mm}
\item   
The normalized distribution in the sum over high$-p_T$ jet energies (dashed)
and in the \gaga\ cms energy \wgg\ (solid) of multiple interaction events with
four jets with rapidity $|y| \leq \ycut = 2$. This value of \ycut\ has been
chosen to approximate the angular coverage of the calorimeter of a typical
TRISTAN detector. Notice that more than 50\% of the events have $\wgg \geq
\sqrt{s}/2 = 29$ GeV.

\end{enumerate}
\end{document}